# Role of Experiments in the Progress of Science: Lessons from our History

D. P. Roy
Homi Bhabha Centre for Science Education,
Tata Institute of Fundamental Research,
V. N. Purav Marg, Mumbai – 400088, India

Abstract

I shall discuss the history of Indian astronomy, Aurveda (life science), chemistry and metallurgy to illustrate how downgrading experiments from scientific learning lead to the decline of ancient Indian science and civilization. We shall see that in the glorious period of ancient Indian civilization, lasting up to the 9th century, there was close interaction between experimental investigations and theoretical analyses in each of these sciences. This was further augmented by two-way interactions with the other advanced civilizations of that time. But both these interactions came to an end around 9th century, leading to the stagnation and decline of Indian science and civilization over the next thousand years. This was the cause rather than the consequence of its subjugation by external invaders, though it was no doubt aggravated by the latter.[1]

**Introduction:**

Let me confess from the beginning that the subject of this article is not the subject of my research. But it is a subject of my concern as an Indian scientist. And I shall present it largely in the words of some Indian scientists of very high esteem, who were deeply concerned about this matter. My only role is one of compilation and occasional elaboration of their comments.

The following sections discuss the ancient Indian chemistry, astronomy, metallurgy and Aurveda in that order. In each case there was close interaction between experimental investigations and theoretical analyses during the glorious period of its history, lasting up to the 9th century. We shall also see that for the three technological sciences of chemistry, metallurgy and Ayurveda, the experimental developments were closely interlinked to one another, so that all the three had a synergetic growth during this period. Moreover, there was a healthy interaction with other advanced civilizations of that time, which particularly influenced the advances in astronomy. However, all these interactions came to an end towards the $9^{th}$ century, leading to the stagnation and decline of Indian science and civilization over the next thousand years. By the $19^{th}$ century the only vestiges of that glorious civilization left was in the form of relics like the Delhi iron pillar and the anecdotal evidences of highly skilled surgery and metallurgy, performed by some illiterate Indian practitioners of these trades. This had profoundly stirred the conscience of the famous scientists of Indian renaissance like Acharya P. C. Ray and Prof. Meghnad Saha, as we shall see below.

---

[1]This article is dedicated to the memory of the famous nuclear physicist and erudite scholar, Prof. Manoj Kumar Pal, who passed away on 3 March 2016.

**Chemistry:**
In his address as the Sectional President of physics and mathematics of the Indian National Science Congress (1926), Meghnad Saha quoted the following lines from a $9^{th}$ century Sanskrit text on chemistry, called 'Rasendra Chintamani' by Dhuduknath, which was brought to his notice by his teacher Acharya P. C. Ray.

'I have heard much from the lips of savants, I have seen many formulae well-established in scriptures, but I am not recording any which I have not done myself. I am fearlessly recording only those that I have carried out before my elders with my own hand. Only they are to be regarded as real teachers who can show by experiments what they teach. They are the deserving pupils, who can actually perform them after having learned from their teachers. The rest are merely stage actors.'

Why was this 9[th] century chemist recording his views on the role of experiments in such strong words? The reason was that by that time the downgrading of experiments from scientific learning and the consequent stagnation of science had already begun in India. Meghnad Saha and P. C. Ray were not only great scientists, but they were also great stalwarts of the Indian renaissance. As such they had a deep understanding of the ancient Indian civilization in its merits as well as its mistakes and limitations. So they were highlighting the latter to the younger generations so that they can learn from these mistakes and overcome the limitations. The definition of stage actors in science was taken quite seriously by Saha at the time of delivering his address; and he issued a warning apprehending that they could vitiate the progress of new science in India [1]. I shall come back to this point and Acharya P. C. Ray's reflections on it at the end of this article. But for now let us continue with the history of Indian chemistry after the 9[th] century.

According to Acharya P. C. Ray [2], Indian chemistry continued to develop for a few centuries after this mainly as the empirical science of alchemy. Alchemy was shunned by Brahmins, but practiced by men of all other castes. There were many pioneers in alchemy; and an outstanding figure named Nagarjuna has been respectfully mentioned in Alberuni's India of early 11[th] century to have lived a century earlier. But there could be several Nagarjunas in history, since Hsuan-tsang in 7[th] century refers to a famous Buddhist alchemist by that name to have lived 5-6 centuries earlier! Alchemy was taught in the monasteries of Nalanda, Vikramasila and Udantapura till their destruction around 1200 AD by Bakhtiar Khilzi. After this the alchemists fled to Tibet and Deccan [3]. P. C. Ray traces back the development of chemistry in India to this subaltern culture of alchemy, which survived through the medieval period, away from the intellectual strata of society [2].

**Astronomy:**
The Calendar Reforms Committee, set up under Meghnad Saha soon after independence, made a thorough review of the three periods of Indian astronomy – i.e. Vedic (→1300 BC), Vedanga (1300 BC – 400 AD) and Siddhanta (400 – 900 AD) periods. According to this review, during the Vedanga period the emphasis had shifted from collecting data from experimental observations to achieving more computational precision. But the Sakas and Kusanas brought the contemporaneous knowledge of Astronomy from Bactria to north-west India. This latest exposure initiated the great spurt of activities towards the end of this period by augmenting the experimentally observed database. This ushered in the Siddhanta era [1]. Surya Siddhanta is assigned to $3^{rd}$ century AD, followed by a quick succession of luminaries : Aryabhatta and Varahamihira (~ 500 AD), Brahmagupta and Bhaskara I (~ 600 AD), Lalla ($8^{th}$ century). Aryabhatta authored Aryabhatiya and a revised version of Surya Siddhanta. He also had a profound influence on the development of Islamic astronomy. So there was a two-way interaction with other cultures during the Siddhanta era. Evidently the interacting cultures all benefited from this, as they could learn from each other's strong points. The following two passages summarize the influence of other cultures on Indian astronomy and that of the Indian astronomy on other cultures [3].

The Yavanajataka was translated from Greek to Sanskrit by Yavaneswara during $2^{nd}$ century AD under Saka king Rudradaman. His capital Ujjain was the "Greenwich of Indian Astronomy". Later in the $6^{th}$ century, Romaka Siddhanta and Paulisa Siddhanta, meaning the treatises of Romans and Paul, were two of the five treatises of Varahamihira called Pancha-Siddhanta. He wrote "The Greeks, though impure, must be honoured since they were trained in sciences and therein excelled others". Similarly Gargi-Samhita says "The Yavanas are Barbarians, yet the science of astronomy originated with them and for this they must be revered like Gods".

These statements illustrate the positive attitude of Indian astronomers to external influence during its glorious era.

On the other hand, Indian astronomy reached China with the expansion of Buddhism during the Han dynasty (25–220 AD). Further translation of Indian works on astronomy was completed in China during the Three Kingdoms era (220-265 AD). However, most detailed incorporation of Indian astronomy occurred only during the Tang dynasty (618-907 AD). Arabs adopted the sine function (inherited from Indian mathematics) instead of chords of arc used in Hellenistic mathematics. Another Indian influence was an approximate formula used for timekeeping by Muslim astronomers. Indian astronomy had an influence on European astronomy via Arabic translations. Muhammad al-Fazari's Great Sindhind, which was based on the Surya Siddhanta and the works of Brahmagupta, was translated into Latin in 1126.

There was a gradual decline in Siddhanta astronomy after the $9^{th}$ century. Although there were great exponents like Bhaskara II ($12^{th}$ century), Nilakantha's Kerala School (15-$16^{th}$ century) and Samanta Chandra Sekhar ($19^{th}$ century), they were few and far between. Let me quote a few lines from the keynote address to a national symposium on Samata Chandra Sekhar by the famous nuclear physicist and ex-director of Saha Institute of Nuclear Physics, Prof. M. K. Pal [1].

'The last exponent of Indian Siddhanta astronomy, Samanta Chandra Sekhar, lived in Orissa from 1835 to 1904. He constructed his own instruments, acquired great skill in using them for accurate observations of sun, moon, planets and stars. When he found by repeated observations that the measured positions in most cases do not agree with results computed using the famous Siddhantas, he boldly concluded that the latter are in error, not his experimental determinations. He wrote his findings in Siddhanta Darpana on palm leaves in Sanskrit using Oriya script. Prof. J. C. Ray of Ravenshaw College, Cuttack, arranged to publish it in Devanagari script through a Calcutta press thirty years later in 1899.'

The most glaring error of the Indian classical Siddhantas is the prediction of the summer and winter solstices (the latter called Makara Samkranti), and the autumn and vernal equinoxes (the latter called Vishuva Samkranti). They were first determined using the simple devise called Gnomon (Sanku in India), in which the direction and length of the shadow of a vertical rod were measured to determine the cardinal directions and time (Fig 1).

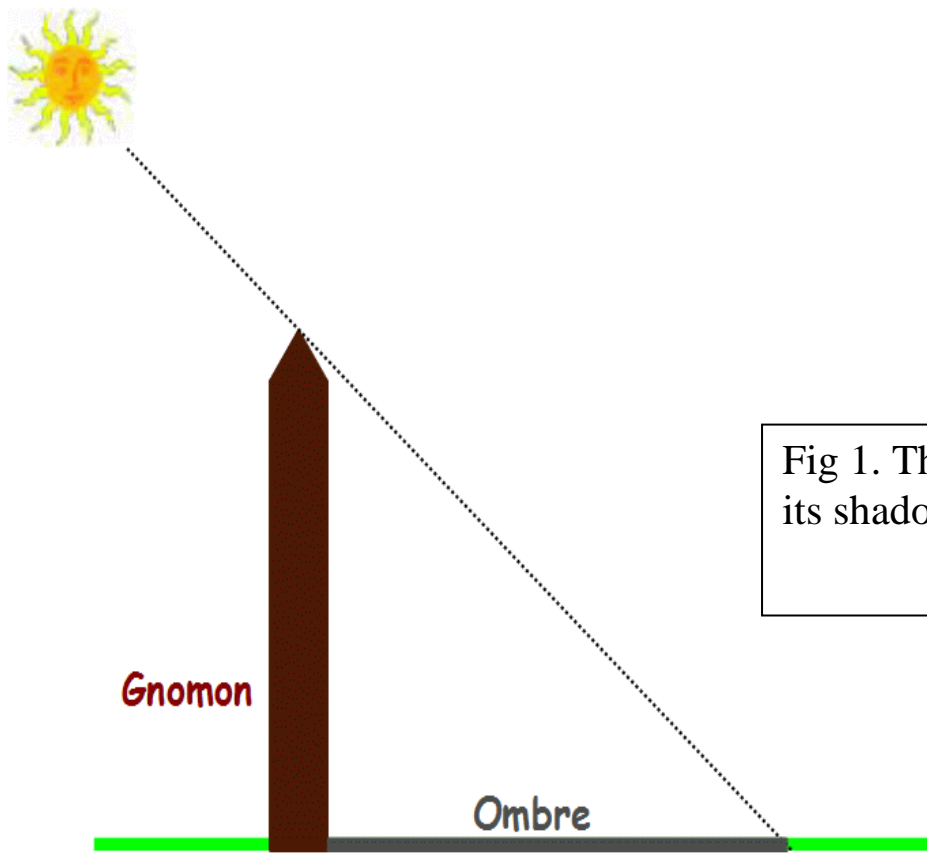

Fig 1. The Gnomon and its shadow [4].

The minimum shadow length marks midday and its direction the cardinal north-south direction. In tropical regions the largest midday shadow length along north (south) marks winter (summer) solstice. And the two mid-points in time between the two solstices mark the two equinoxes. At the time of this calibration around 400 AD Helial (Sun synchronous) rising of the constellation Capricorn (Makara) matched with the winter solstice of 21-22 December and that of Cancer (Karkata) matched with the summer solstice of 21-22 June. That is why the southern and northern tropics were named tropics of Capricorn (Makara kranti) and Cancer (Karkata kranti), while the equinoxes matched with the vertical alignment of sun over the equator (Vishuva rekha). However, precession of the earth's rotation axis over the past 1600 years has resulted in a 23 days gap between the celestial and terrestrial markers. Evidently the terrestrial events like the change of season and harvest of crops are determined by the true solstice and equinox times corresponding to the terrestrial markers rather than the celestial ones. This is a glaring example of how blind following of the ancient scriptures without experimental recalibration leads to wrong solstice and equinox times.

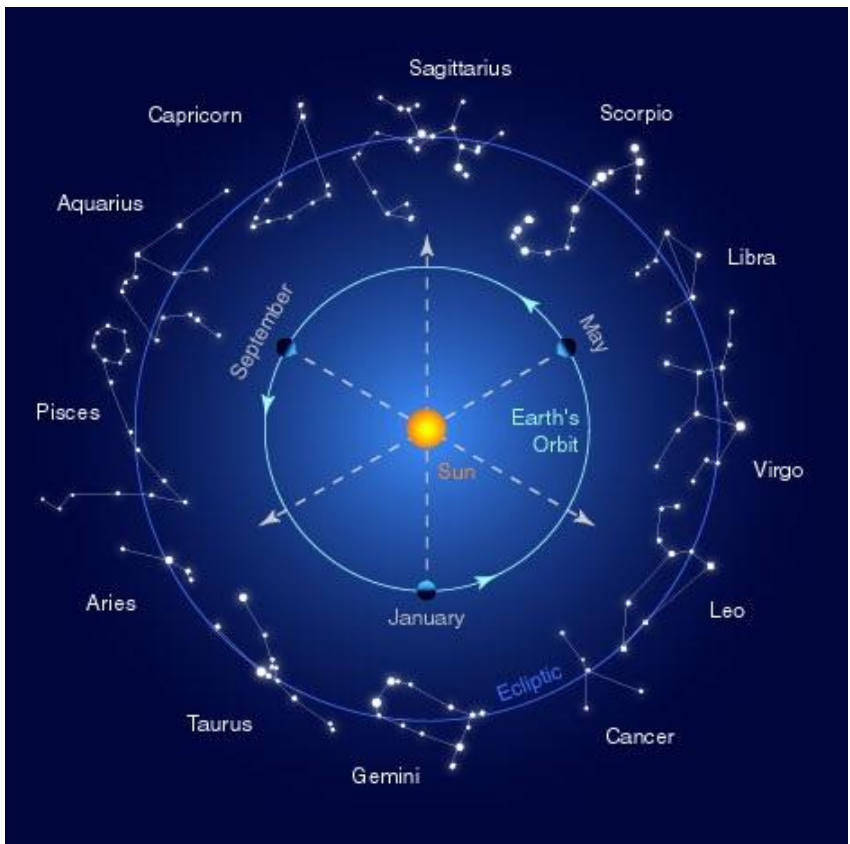

Fig 2. Present solar alignment of different constellations over the year [5].

Fig 2 [5] is taken from an article of Prof. M. N. Vahia on why we observe Makara Samkranti on 14 January [6]. It clearly shows that the present solar alignment with the constellation Capricorn (Makara) indeed starts at mid-January instead of the true winter solstice of 21-22 December. The slow time drift of the solstice and equinox was empirically known to the ancient Greek astronomers. Therefore it must have been known to the Siddhanta astronomers of India as well. So the question is why the necessary recalibration to account for this drift was not done. The reason could be one of societal attitude. Firstly to dirty your hands with experiments; and secondly when you find after yearlong painstaking observations that your empirical results are in conflict with the predictions of time honoured scriptures, who will listen to you? So the astronomers by and large chose the easy option of following the scriptures on the excuse that the Makara Samkranti corresponds to the alignment of sun with the celestial Makara constellation rather than the terrestrial Makara kranti, although the former has little relevance to the terrestrial phenomena as mentioned above. In many parts of India the Vishuva Samkranti on 14 April, marking the start of the solar month of Baishakh, is even a more important festival than the Makara Samkranti. It marks the start of the new year in Bengal and Orissa, and also in Kerala, where it is called Vishu Samkranti. It is celebrated in Mangalore as Bishu and in Assam as Bihu. It again comes after 23 days of the true Vernal Equinox; and in this case one does not even have the alibi of a celestial marker by that name. The Indian Calendar Reforms Committee had suggested removing the historical misnomers from the Baishakh Samkranti of 14 April and Magh Samkranti of 14 January, and recognize the true Vernal Equinox and Winter Solstice in the Indian calendar as Vishuva rekha Divas and Makara kranti Divas respectively. But it went unheeded.

Another serious limitation of Indian astronomy of this period is the non-recording of purely empirical phenomena. The Chinese have kept data of meteoric showers, 29 appearances of Halley's comet, 90 novae and supernovae along with intense sunspot activities [1]. Yet there is no Indian record of these empirical phenomena, presumably because they did not relate to astronomical theories of that time. In particular, the spectacular Crab supernova explosion of $11^{th}$ century appeared as the second brightest object after the moon in the night sky for several weeks. It has been recorded by the Chinese, Arab and even Mayan astronomers of Mexico. Yet there is no credible evidence of Indian astronomical record of this very important event. This was the conclusion of Profs. J. V. Narlikar and S. Bhate after a thorough search of the contemporary Indian documents on an INSA project.

**Metallurgy [3]:**
India was a major exporter of ferrous metals throughout ancient history. The iron pillars of Delhi, originally from Vididsha (400 AD), and of Dhar (1000 AD) stand living testimony to the skills of ancient Indian metallurgists.

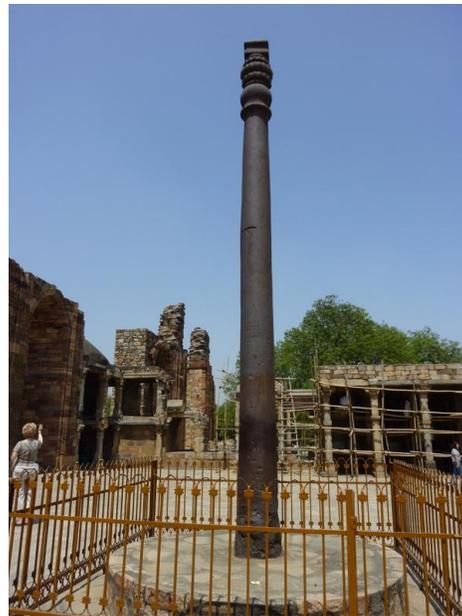

Fig 3. The iron pillar of Delhi

The Delhi pillar is 7 m high and weighs 6.5 tons. It is 98% pure iron with a high Phosphorous content to make it rust resistant. It is generally believed that no other country had the capability to produce an iron mass of this size and purity till the industrial revolution of 18$^{th}$ century. The Dhar pillar had a weight of a little more than 7 tons and almost twice the height of the Delhi pillar, but is now broken into three pieces. It also has a high Phosphorous content for rust resistance like the Delhi pillar [7].

Equally important was the discovery of steel production in Deccan by the carbonization of iron around 600 BC [3]. It was globally exported throughout the period of ancient Indian history. There was a close triangular link between Alchemy, Metallurgy and Aurveda. Alchemy had two branches called Deha Siddhi and Loha Siddhi. The former dealt with the production of various Bhasmas of Aurvedic medicine, while the latter dealt with the chemicals used in metal smelting and production of special quality metals like steel. The latter in turn was closely connected with the sharp edged instruments used in Aurvedic surgery. It is said that the surgical instruments of Susruta were fabricated with Deccan steel. These three sciences had a synergetic growth through the period of ancient history up to the 9$^{th}$ century.

The state of Indian metallurgy after 1000 AD has been discussed by Prof. B. Prakash [8]. It saw a rapid decline during 11$^{th}$-12$^{th}$ century as Ghaznavid and Ghorian invaders destroyed the iron producing industry and took away many thousands of skilled workers as slaves to bolster their own armament production. However, during the Mughal period a subaltern culture of metallurgy was revived for large scale production of armaments and construction of large cannons, some weighing 20-40 tons. With some interruptions the Deccan steel export to Arab countries continued into the medieval period for making quality armaments. The famous Damascus sword was made with Deccan steel [3]. But both of them had declined by the 18$^{th}$ century. The death blow to the Indian metal industry was dealt by the British Colonial Government policy of shipping iron ore to British plants at the cost of the Indian foundries.

**Ayurveda:**
The Indian Academy of Sciences has brought out a vision document on Aurvedic Biology by Prof. M. S. Valiathan, who is a prominent cardiac surgeon and discoverer of a famous heart valve, past president of Indian National Science Academy, past vice-chancellor of Manipal University and presently a national professor there. Prof. Valiathan is also an authority on Ayurvedic Biology not only as a theoretical scholar but one actively engaged in experiments to scientifically test the efficacy of various Ayurvedic procedures as well. Therefore my discussion of this section will be largely based on the material of this document [9]. In fact I shall be quoting many passages from this document, risking the charge of plagiarism, because I could not have put them any better.

According to Valiathan [9], the Samhita phase from $1^{st}$ to $9^{th}$ century AD is regarded as the golden age of Ayurveda. It had three major texts called the Brihadtrayi. 1) Caraka Samhita ($1^{st}$ century) is a redaction by Caraka of a treatise composed by Agnives several centuries earlier. 2) On the other hand, Susruta Samhita ($2^{nd}$-$3^{rd}$ century) is a redaction by Nagarjuna of the surgical treatise of Susruta, who is said to have lived around 700 BC. 3) Finally, Astanga Samgraha and Astanga Hrdaya ($8^{th}$–$9^{th}$ century) are composed by Vagbhata.

Caraka's redaction was so highly creative that the new text came to be acclaimed as Caraka Samhita. Here Ayurveda got its name for the first time, and it moved from a faith-based to a reason-based platform. It was encyclopedic in the coverage of medicines, and recognized as the last word in internal medicine. It was translated into Persian, Arabic and Tibetan within 2-3 centuries and spread its influence to central Asia, where Bower manuscript of 400 AD with numerous quotes from Caraka was discovered in 1890. Bower was an intelligence officer of British Indian army, who discovered this manuscript written on Parchment in Brahmi script with some natives of central Asia and brought it to the Asiatic Society at Calcutta [3]. Caraka Samhita was translated into English in 19$^{th}$ century. Its popularity continues in the 21$^{st}$ century, when a digital version was prepared by Prof. Yamashita of Kyoto University [9].

Susruta's name is forever associated with rhinoplasty (nose repair), the only surgical procedure from India to have won global recognition in three millennia! Susruta Samhita is a comprehensive medical treatise with heavy surgical orientation, dealing with surgical procedures, instruments, care of trauma, medications etc. Drawings of some surgical procedures and instruments are shown in Figs 4-7 [9]. Compared to Caraka Samhita it has simpler language and lower emphasis on the philosophical dimensions of medical practice. This, along with its precise drawings of surgical procedures and instruments, suggest the compiler Nagarjuna was more likely to be a hands-on experimentalist rather than a theoretical scholar. Many believe him to be the Buddhist alchemist of that name described by Hsuan-tsang; but there is no definitive evidence for this [3]. Susruta Samhita enjoyed great authority even beyond the Indian borders because it was translated into Arabic under the Caliphate, when Indian physicians were believed to have lived in Baghdad [9].

There is little doubt that the Susruta and Carak Samhitas were taught at Nalanda; and the large number of students from Tibet, China and other east Asian countries would have carried home their copies and translations. Transfer of knowledge was also facilitated by Indian teachers accompanying these home-bound disciples. Even today, several texts in medicine, philosophy etc, which are no longer available in Sanskrit original, are available in their Chinese and Tibetan translations. What the barbarians destroyed in India had a resurrection in other countries [9]. The last sentence refers to the destruction of Nalanda by Bakhtiar Khilzi around 1200 AD.

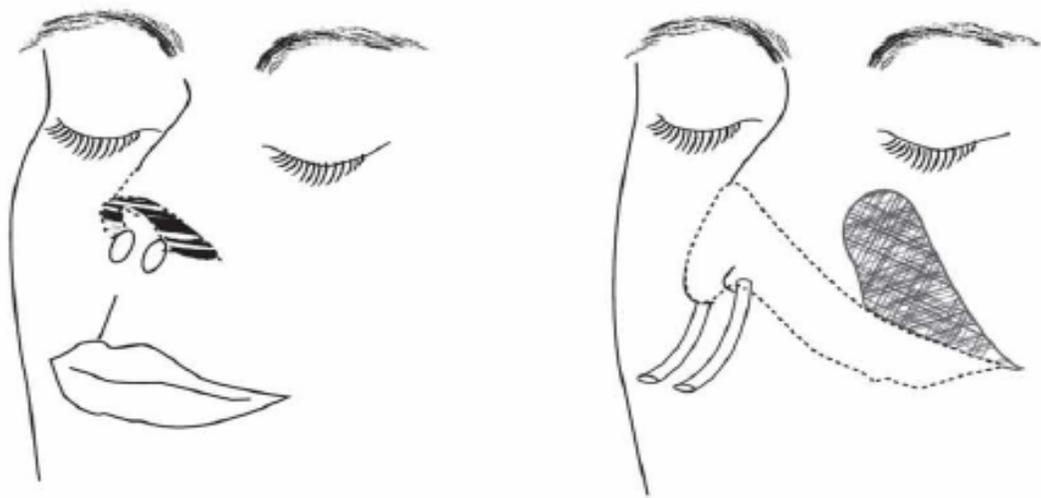

Fig. 4 Plastic repair of nose
Described by Susruta: a pedicle flap from the cheek was used; the eighteenth century practitioner in Pune took the flap from forehead

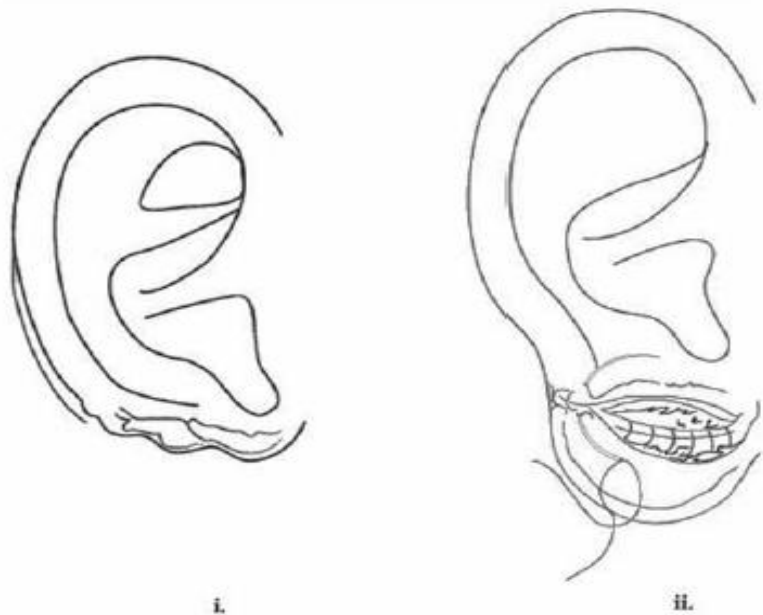

Fig. 5 Plastic repair of ear lobe
was recommended by Susruta when the ear lobe was destroyed by infection following the piercing of ear

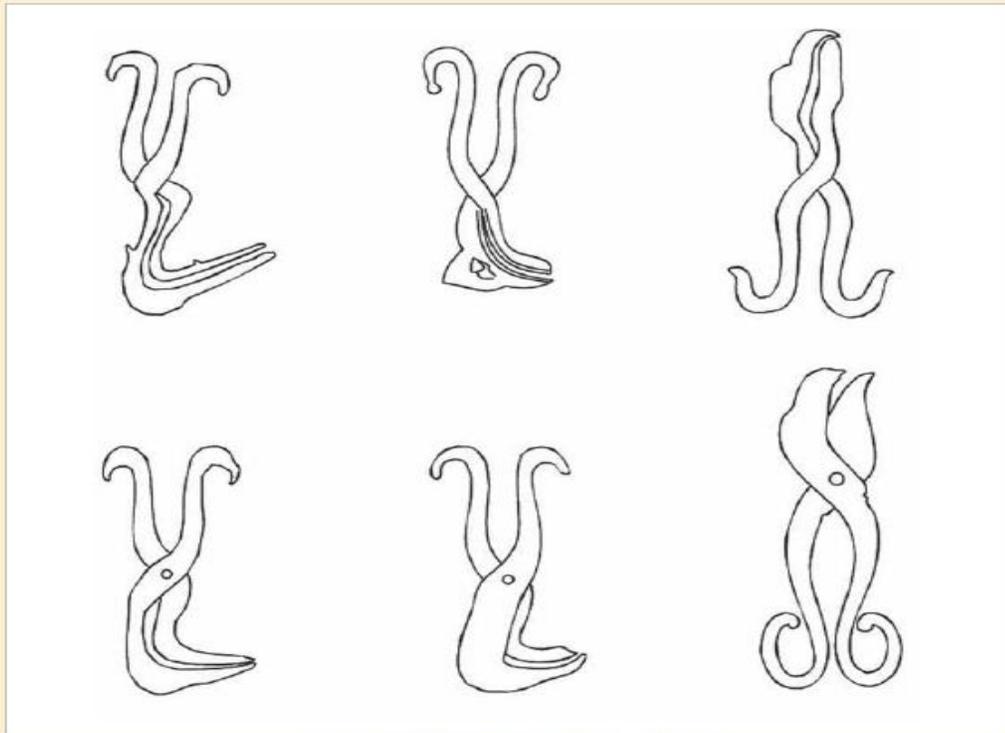
Fig. 6 Instruments – blunt (Yantras)
A few from the 100 blunt instruments of Susruta

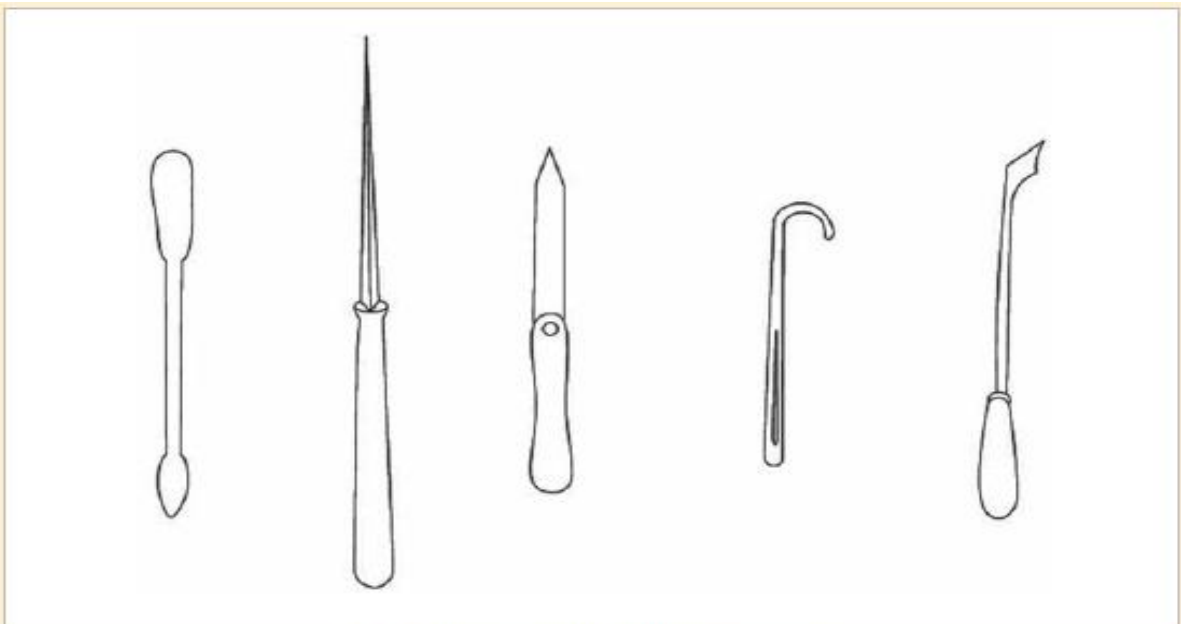
Fig. 7 Instruments – sharp (Sastras)
A few from the 20 sharp instruments of Susruta

Acharya P. C. Ray estimated the date of composition of Astanga Samgraha and Astanga Hrdaya by Vagbhata to be $8^{th}$-$9^{th}$ century, when Ayurveda was on the threshold of stagnation [2]. These texts accept the authority of Caraka and Susruta in no uncertain terms and present their teachings in a simple and abridged manner for average students. Astanga Hrdaya accomplished this objective admirably and became a popular favourite, thanks to the gift of poetic excellence that no other text could claim. After Vagbhata, the springs of creativity ran dry and a long phase of stagnation ensued for a thousand years in the history of Ayurveda [9].

Of course some important texts of Ayurveda appeared during this phase along with many dictionaries and commentaries on earlier texts. But there were no more Carakas and Susrutas, nor the power-houses of learning like Nalanda. The preference of the Muslim rulers for Unani hastened the decline of Ayurveda. But the malady had roots running deeper in the social history of India, because the surgical techniques of Susruta had more or less disappeared from the mainstream of Ayurveda already by the time of Vagbhata. Cadaveric dissection was no more mentioned; and the training of disciples did not include exercises on cucumber, jackfruit and animal skin etc for learning incision, extraction, scraping and other surgical procedures [9].

So the Afghan/Turkish conquest of India and destruction of Nalanda around 1200 AD were not the causes but rather the consequences of the decline in Indian science and civilization that had started at least a couple of centuries earlier. Mahmud of Ghazni raided India 17 times during 1000-1027 AD over a wide front from Mathura to Somnath in Saurastra, destroying its monuments and industries, plundering its wealth and taking many thousands of its skilled workers as slaves. Yet we did not learn our lessons and put our house in order. Al-Beruni was a central Asian scientist/scholar, who came to India in 1017 at the behest of Mahmud of Ghazni and spent thirteen years travelling through this country to write a comprehensive book on the nation and its people. His account is generally considered to be candid but objective. An extract from Al-Beruni's account of the Indian people is quoted below [10].

'The Hindus believe that there is no country but theirs, no nation like theirs, no king like theirs, no religion like theirs. They are haughty, foolishly vain, self-conceited and stolid. They are by nature niggardly in communicating that which they know, and they take greatest possible care to withhold it from men of another caste among their own people, still much more, of course, from any foreigner….. Their haughtiness is such that, if you tell them of any science or scholar in Khorasan or Persia, they will think you to be both an ignoramus and a liar. If they traveled and mixed with other nations, they would soon change their mind, for their ancestors were not as narrow minded as the present generation is.'

The last line of this passage is very significant, because the nation had assimilated the Saka and Kusana conquerors into the Indian civilization in $2^{nd}$-$3^{rd}$ century AD. It had also spread the Indian civilization throughout south-east Asia through travelling tradesmen without any bloodshed. And it had spread Buddhism over most of Asia through exchange of scholars. But this vibrant nation with a pan-Asian outreach had folded up into its narrow regional, caste and sub-caste groups by 1000 AD. Moreover, lofty institutions like Nalanda had weakened considerably, because there was no empire to support them anymore. So it became an easy prey for external aggressors.

**Surviving Subcultures of Surgical and Metallurgical Skills:**
The surgical procedures which disappeared from the main stream of society survived however among castes, considered low in the social hierarchy. Susruta's nose repair is an interesting example. Barring a perfunctory reference, it received no serious attention in the Aurvedic texts; nor was it performed by reputed Vaidyas. Its survival was "discovered" accidentally by British observers in Pune towards the end of 18$^{th}$ century [9].

Pune Nose Repair Episode: Dr. Scott, a sympathetic British doctor residing in Mumbai, had heard from one Capt. Irvine in 1793 about the practice among "gentoos of putting new noses on people who have had them cut off" presumably for some criminal offence. He assured Dr. Scott that all the employees of the East India company in Pune were witness to the operation which gave them a "pretty good nose"! Dr. Scott then wrote to Mr. Findlay, the company surgeon in Pune, to ascertain the veracity of this report because such an operation was unknown in Europe. Mr. Findlay sent a detailed report on the basis of eyewitness observation by himself and Mr. Cruso on 1$^{st}$ January 1794. The report described how a "koomar" caste man had borrowed an old razor for this occasion, dissected a flap from the forehead of the patient with much composure, freshened the edges of the nasal defect and applied the flap there on by rotation with a cement "without the aid of stitches, sticking plaster or bandages". The flap healed and "an adhesion had taken place seemingly in every part". It was a report of this procedure, published in the "Gentleman's Magazine" of London in 1794, which caught the attention of a surgeon, Dr. J. C. Caprue, FRS. He performed the operation for the first time in the West and published a full length paper on "An account of two successful operation for restoring a lost nose from the integuments of the forehead" in 1816.

Other Surgical Skills: A similar eyewitness report on Susruta's couching for cataract was given by Dr. Ekambaram of Coimbatore in 1916. He found that the procedure was done by iterant Mohammedan Vaidyas who followed the steps of Susruta's method [9]. Note that the procedures in Pune and Coimbatore were done not by Ayurvedic physicians but by illiterate men, who had learned the techniques from an earlier generation. Treatment of fracture by bonesetters, child delivery by dais and many other procedures involving "dirtying of hand" were relegated to lower caste persons, who did not understand their anatomical basis or rationale. It was as if the nation's brain had been decoupled from its hand, which ensured that there could never be innovation based on true understanding.

Metallurgical Skills: There is also anecdotal evidence showing the survival of a subculture of metallurgical skills among the lower castes [9, 11]. On the request of the Govt. of Bengal in 1828, James Franklin, FRS, made a thorough study of the ore, charcoal and furnaces used by the natives of Central India for making iron. He wrote "the smelting furnaces, though crude in appearance, are nevertheless very exact in the interior proportions, and it has often surprised me to see men, who are unquestionably ignorant of their principle, construct them with such precision". He went on to describe in detail the geometrical and practical construction of the furnace, the construction and use of bellows, construction of two refineries for each furnace, mode of smelting and refining etc. On getting the product evaluated at the Sagar mint he wrote " the bar iron was of the most excellent quality, possessing all the desirable properties of malleability, ductility at different temperatures and of tenacity of which I think it cannot be surpassed by the best Swedish iron". Though the workmen could not answer Franklin's questions or explain the procedures used for hundreds of years by their forefathers, he commented that the "original plan of this singular furnace must have been the work of advanced intelligence" [11]. In fact this was the relic of a civilization that had produced the iron pillars of Delhi/Vidisha in 400 AD and Dhar in 1000 AD. Actually Vidisha and Dhar are both located in Central India, i.e. the same geographical region as the abovementioned workmen of a much later period.

**Conclusion:**
The above anecdotes make poignant stories. But what lesson do we learn from them? Should they make us happy or sad? Let me conclude by answering these questions in the words of Prof. Valiathan and those of his inspiration, Acharya P. C. Ray, as quoted by him.

Reflections of M. S. Valiathan: The workmen doing the nose repair in Pune, cataract couching in Coimbatore and ore smelting in Jabalpur were condemned to illiteracy, low social status, poor self-esteem and little hope of self advancement. Since this grim prospect claimed hundreds of thousands of citizens, who used their hands to make a living, ruin could be the only destination of their nation [9].

Reflections of P. C. Ray: According to Susruta, the dissection of dead bodies is a sine qua non (indispensible) to the students of surgery, and this high authority lays particular stress on knowledge gained from experiments and observations. But Manu would have none of it. According to Manu, the very touch of corpse is enough to contaminate the sacred person of a Brahmin. Thus we find shortly after Vagbhata, the handling of a lancet was discouraged and anatomy and surgery fell to disuse and became, to all intents and purposes, lost sciences for the Hindus. It was considered equally undignified to sweat away at the metal furnaces. The sciences being thus relegated to the lower castes, and the professions made hereditary, a certain degree of fineness, delicacy and deftness in manipulation was no doubt secured. But this was accomplished at a terrible cost. The intellectual portion of the community being thus withdrawn from active participation in these sciences, the how and why of phenomenon – the coordination of cause and effect – were lost sight of. The spirit of enquiry gradually died out among a nation, naturally prone to speculation and metaphysical subtleties, and India for once bade adieu to experimental and inductive sciences. Her soil was made morally unfit for the birth of a Boyle, a Descartes, or a Newton; and her very name was expunged from the map of the scientific world for a time [2].

Under these circumstances, India's rout at the East-West encounter of the 18$^{th}$ century was a foregone conclusion [9].

**Acknowledgement:** The author acknowledges partial support from the senior scientist fellowship of the Indian National Science Academy.